\input epsf
\documentstyle[12pt]{article}

	\newcommand{\be}{\begin{equation}}
	\newcommand{\ee}{\end{equation}}
	\newcommand{\ba}{\begin{eqnarray}}
	\newcommand{\ea}{\end{eqnarray}}
	\newcommand{\ban}{\begin{eqnarray*}}
	\newcommand{\ean}{\end{eqnarray*}}
	\newcommand{\barr}{\begin{array}}
	\newcommand{\earr}{\end{array}}

	\newcommand{\et}{\hspace{-0.08in}{\bf .}\hspace{0.1in}}

\def\nn{\nonumber}
\def\bs{\bigskip}

\def\qq{\qquad}

\def\r{\rho}
\def\a{\alpha}
\def\b{\beta}

\def\d{\delta}

\def\e{\epsilon}

\def\m{\mu}
\def\n{\nu}
\def\na{\nabla}
\def\om{\omega}

\def\pa{\partial}

\def\frac#1#2{{#1\over#2}}

	\newtheorem{ex}{Example}
	\newcommand{\bex}{\begin{ex}\et}
	\newcommand{\eex}{\end{ex}}

	\newcommand{\ed}{\end{document}}
\def\jcal{{\cal J}}

	\def\sqr#1#2{{\vcenter{\vbox{\hrule height.#2pt
		\hbox{\vrule width.#2pt height#1pt \kern#1pt
		 \vrule width.#2pt}
		\hrule height.#2pt}}}}

\def\x{{\vec x}}
\def\y{{\vec y}}

\newcounter{figg}

\begin{document}
\baselineskip 15 pt

\vspace{1cm}
	\begin{center}
	{\Large 3-Cocycles and the \\
	Operator Product Expansion \\ }
	\vspace{1cm}
	Javier P.\ Muniain$\,{}^{\dagger}$ and
	Jos\'e Wudka$\,{}^{\ddagger}$ \\
	\vspace{1cm}
	{\small{\it Department of Physics \\
	University of California, Riverside \\
	California 92521-0413, U.\ S.\ A. \\ }}
	\vspace{1cm}
	{\small December 19, 1995 \\}
	\vspace{1cm}
	\end{center}

\begin{abstract}
Anomalous contributions to the Jacobi identity of chromo-electric
fields and non-Abelian vector currents are calculated using a non-perturbative
approach that combines operator product expansion and a generalization of
Bjorken-Johnson-Low limit. The failure of the Jacobi identity and the
associated 3-cocycles are discussed.
\end{abstract}

\vspace{3cm}
{\footnotesize ${}^{\dagger}$muniain@phyun0.ucr.edu \qq
${}^{\ddagger}$jose.wudka@ucr.edu \\ }
\vfill\eject

\begin{center}
\subsection*{1. Introduction}
\end{center}

The study and evaluation of commutators, as well as their algebraic properties
has been the motive of much research over the past
years. Many results, leading to important measurable effects
were found using canonical commutation relations, which, unfortunately,
are often ill defined. This was made clear
by Schwinger \cite{Schwinger} in his
evaluation of the matrix element
$\langle 0|[J_0(x),J_i(y)]|0\rangle$
($ J^\mu $ denotes a current)
at equal time. This commutator has a
non-canonical term proportional to the gradient of a delta function
which is mandated by locality, Lorentz covariance, positivity and
current conservation, and which is not generated following (naive)
canonical manipulations.

Much work has been done towards finding perturbative expressions for the
commutators \cite{pert}. Recently, an effort
was made to find a practical method to evaluate commutators in
a non-perturbative way \cite{mw1} based on the operator product
expansion (OPE) \cite{Wilson} and on the Bjorken, Johnson and Low (BJL)
 \cite{BJL}
definition of the commutator. The BJL definition preserves all desirable
features of the theory, and reproduces the canonical results whenever
these are well defined \cite{Jackiwbook}.
In the present paper we generalize the method proposed in \cite{mw1}
to the case of double commutators~\footnote{For a related
publication see Ref. \cite{doublecom}.}, in particular we will study violations
of the Jacobi identity.
The present approach is based on a double high-energy
limit (taken in a particular order) of the Green function for three
local operators.

Given any three operators $A$, $B$ and $C$ we define the quantity
\be
{\cal J}[A,B,C]= [[A,B],C] + [[B,C],A] + [[C,A],B].
\label{jacobi}
\ee
which vanishes whenever the Jacobi identity is preserved.
Before we proceed it is worth pointing out that in a theory where all
the linear operators are well defined no violations of the Jacobi
identity can appear, and $ {\cal J } $ is identically zero.
In this paper we will consider models in which
the operators and their products require regularization, for such theories we
construct an operator which is naively equal to $ \cal J $
(that is, it coincides with the expression (\ref{jacobi}) whenever the
operator products are well defined), but which has
finite matrix elements and respects all the desirable symmetries of the model.
The  price is that not all such matrix elements need vanish. The
procedure we describe below provides a {\em definition} of $ \cal J $.

Situations in which $ {\cal J} \not= 0 $ present problems in
providing well-defined representations for the corresponding algebra of
operators.
A non vanishing $ \cal J $ is then understood as an obstruction in constructing
such representations in terms of operator-valued
distributions~\cite{carey.et.al.}.
However, objects which are local in time and obey $ { \cal J } \not= 0 $ may
still be
defined in terms of their commutators with space-time smoothed operators.

The expression we obtain for $ \cal J $ depends on a small number of
undetermined
constants. The present method is not powerful enough to determine
whether such constants are non-zero. Nonetheless it is still possible to
obtain some non-trivial information concerning the expression for our
definition of $ \cal J $ mainly based on the consistent implementation
of the model's symmetries. We will comment on this fact in the last
section.

It is well known \cite{Jackiw2,3cocy1,magmon}
that violations of the Jacobi identity $ \jcal = 0 $
within an algebra generate, in general,
violations of associativity in the corresponding group.
If the group generators, denoted by $ G_a $, satisfy
\be
{\cal J} \left[ G_{a_1},G_{a_2} ,G_{a_3} \right] =
{i\over{3!}}\om_{[a_1a_2a_3]} \not= 0
\ee
($[a_1a_2a_3]$ denotes antisymmetrization in all variables $a_i$)
the corresponding
lack of associativity is parameterized by the three-cocycle $ \omega
$ (for a review see Refs. \cite{Jackiw2}).
Consistency requires the closure relation \cite{3cocy1}
\be
f_{ c [ a_1 a_2 } \omega_{ a_3 a_4 ] c } = 0 \label{closurerel}
\ee
(where summation over $c$ is understood).

The existence and properties of 3-cocycles has been
under investigation in quantum field
theory for some time now.
The behavior of gauge transformations in an anomalous gauge theory,
as well as in a consistent gauge theory with Chern-Simons term, can be given
a unified description in terms of cocycles \cite{chernsimons}.
Violations of the Jacobi identity also appear in the quark model: if
the Schwinger term in the commutator between time and space components
of a current is a c-number, the Jacobi identity for triple commutators
of spatial current components must fail \cite{3cocy2}. This fact has
been verified in perturbative BJL calculations \cite{3cocy3}{

In the context of quantum mechanics, 3-cocycles
appear in the presence of magnetic monopoles \cite{magmon}.
For example, a single particle moving in a magnetic field $ \vec B$
satisfies $J[v^1, v^2, v^3]= (e \hbar^2/ m^3)\vec\nabla
\cdot \vec B$, where $v^i$ represent the components of the (gauge
invariant) velocity operator. If $\vec\na\cdot\vec B\not=0$,
as in the case of a point monopole, the Jacobi identity fails.

The paper is organized in the following manner. Section 2 is
dedicated to the description of the method. Section 3, as an
application of the method, studies the
failure in field theory of the Jacobi identity for chromo-electric fields.
Following this, Section 4 is dedicated to 3-cocycles associated to
the QCD quark charges and Gauss' law generators. The results of these
sections are compared to the results derived form perturbation theory in
section 5. Conclusions are presented in Section 6.

\begin{center}
\subsection*{2. Description of the method}
\end{center}

In this section we will generalize the method proposed in \cite{mw1}
to study double commutators and the possibility of violation of Jacobi
identity. The canonical
evaluation of equal time commutators sometimes presents ambiguities
 \cite{Schwinger}, and it becomes necessary to have an alternative way
to define and calculate these objects. This is achieved by the
Bjorken, Johnson and Low \cite{BJL} {\it definition} of the single
commutators (for a review see
 \cite{Jackiwbook}) which relies
only on the construction of the time-ordered product of the operators
whose commutator is desired. Specifically, the commutator of $A$ and $B$
is obtained from
\ba && \lim_{p^0 \to \infty} p^0 \int d^n x \;e^{i p x} \langle \a |
T A(x/2) B(- x/2) | \b \rangle = \nn \\
&& \qq \qq \qq = i \int d^{n-1} x\; e^{-i {\vec p}\cdot \x} \langle\a |
[A(0,\x/2), B(0,-\x/2)] | \b \rangle. \label{1}
\ea
where $p^0$ stands for the time component of the four-momentum.
The BJL definition (\ref{1}) uses the time ordered
product $T$, which (in general)
is not a Lorentz covariant object \cite{Jackiwbook}, while in field theory
(e.g. Feynman diagrams in perturbation theory) one calculates
an associated covariant object, usually denoted by $T^*$. The difference
between $T$ and $T^*$ is local in time,
involving $ \delta( x_0) $ and its derivatives
 \cite{Jackiwbook}, which translates into a
polynomial in $p^0$ in momentum space. Therefore
in (\ref{1}) we can replace $T$ by $T^*$ provided we
drop all polynomials in $p^0$. Equivalently, the
Fourier transform of the commutator is the residue of
the $1/{p^0}$ term in a Laurent expansion of the time ordered
product $T^*$ (divided by $i$).

This approach can easily be extended to the study of
double commutators. We first define
\be
{\cal C}(p,q) = \int d^n x d^n y \, e^{i (p x + q y)} \langle \a | T A(x)
B(y) C(0) |\b \rangle, \label{green}
\ee
and use the (formal) identities
\ba
\frac{\pa}{\pa x_0} T A(x) B(y) C(0) &=& T(\dot A B C) +
\d(x_0 - y_0) T([A,B]_{(x_0)} C(0)) \nn \\
&+& \d( x_0 ) T(B(y_0)[A,C]_{(x_0)}) \\
\frac{\pa}{\pa y_0} T A(x) B(y) C(0) &=& T(A \dot B C) +
\d(y_0 - x_0) T([B,A]_{(y_0)} C(0)) \nn \\
&+& \d( y_0 ) T(A(x_0)[B,C]_{(y_0)}),
\ea
where the subscript in the commutators
indicates the common time of the operators.

To simplify the resulting expressions we define
\font\smallii=cmr8 scaled\magstep2
\def\blim#1#2{{}_{#1}\hbox{{\smallii I}}\!\hbox{{\smallii L}}_{#2}}
\be
\blim qp = \lim_{q_0 \to \infty } q_0 \quad
 \lim_{p_0 \to \infty; q_0 = {\rm const}} p_0 \label{blimdef}
\ee
and obtain, after straight-forward manipulations,
\ba
\blim qp {\cal C}(p,q) &=&
\int d^{n-1}x\; d^{n-1}y\; e^{-i(\vec p \cdot \x + \vec q \cdot \y)}
\langle \a |[ B(0,\y) , [ C(0) , A(0,\x) ] ]| \b \rangle \nn \\
\blim pk {\cal C}(p,-p-k) &=&
\int d^{n-1}x\; d^{n-1}y \;e^{-i(\vec p \cdot \x + \vec q \cdot \y)}
\langle \a |[ A(0,\x), [ B(0,\y),C(0)] ]| \b \rangle \nn \\
\blim kq {\cal C}(-q-k,q) &=&
\int d^{n-1}x\; d^{n-1}y\; e^{-i(\vec p \cdot \x + \vec q \cdot \y)}
\langle \a |[ C(0), [ A(0,\x), B(0,\y)] ]| \b \rangle, \label{2com}
\ea
where $ k = - p - q $. These expressions imply
\be
\left( \blim qp + \blim pk + \blim kq \right) {\cal C} = \int d^{n-1} x
d^{n-1} y e^{ - i \vec p \cdot \x - i \vec q \cdot \y } {\cal J} [ A , B , C ]
\label{jidef}
\ee
where, as above, $ { \cal J } [ A , B , C ] = [ A , [ B , C ]] +
[ B , [ C , A ]] + [ C , [ A , B ]] $.

The above manipulations suggest that we {\it define} $ {\cal J} [ A , B , C ]
$ via (\ref{jidef})~\footnote{When canonical manipulations are well
defined we will have $ {\cal J} [ A , B , C ] = 0 $.}.
In the following we will use this definition of $ { \cal J} $

Since we are interested in the large-momentum-transfer behavior, it
is appropriate to express the product of operators in (\ref{green}) as
a sum of non-singular local operators with possibly singular c-number
coefficients \cite{Wilson},
\be
\int d^n x d^n y\;e^{i (p x+ q y)} \langle \a |T A(x) B(y) C(0)| \b
\rangle = \sum_i c_i(p,q)\; \langle \a | {\cal O}_i (0)|\b \rangle,
\label{2}
\ee
each term in the
OPE should respect the same symmetries (and possess the same
internal quantum numbers) as the
Green's function (\ref{green}). As for the single commutator case
 \cite{mw1} it is more convenient to derive the various (double)
commutators from the covariant time-ordered product $T^*$. the
difference $ T[ A(x) B(y) C(z) ] - T^* [ A(x) B(y) C(z) ] $ is an operator
local in $ x-y $ or $ y-z $ or $ x-z $. Thus we will drop all terms
proportional to a polynomial in $p^0$, $q^0$ or $ k^0$.

Substituting
(\ref{2})
in (\ref{jidef}) we obtain
\ba
&&\int d^{n-1}x\; d^{n-1}y\; e^{-i(\vec p \cdot \x + \vec q \cdot \y)}
\langle \a |{\cal J}[A(0,\x), B(0,\y), C(0)] | \b \rangle = \nn \\
&& \mskip 200 mu = \sum_i \left( \blim qp + \blim pk + \blim kq \right)
c_i(p,q)\langle \a |{\cal O}_i (0)|\b \rangle
\label{2ope}
\ea

It is worth pointing out that similar manipulations have been used to
provide constraints on the general form of current anomalies \cite{old}.

\begin{center}
\subsection*{3. Jacobi Identity for chromo-electric fields}
\end{center}

In this section we investigate the existence of 3-cocycles associated
with the (chromo) electric fields of a gauge theory,
denoted by $E^a_i = F^a_{0i}$, where $F^a_{\m\n}$ is the non-Abelian
gauge field strength. We will consider the four-dimensional case first
and then briefly consider the case of two dimensions.

We evaluate the Jacobi operator for three chromo-electric fields
by studying the behavior of the correlator of three strength tensors
$T \left\{ F_{\m_1\n_1}^{a_1}(x_1) F_{\m_2\n_2}^{a_2}(x_2)
F_{\m_3\n_3}^{a_3}(x_3) \right\} $ for the case of $\m_r=0$, $\n_r \not= 0$.

Following (\ref{green}) we consider
\be
{\cal C}^{a_1 a_2 a_3 }_{ \m_1 \n_1 \m_2 \n_2 \m_3 \n_3 }(k_1,k_2)
= \int d^4  x_1 d^4 x_2 \, e^{i (k_1 x_1 + k_2 x_2)} \left\langle \a \left|
T^* \left\{ F_{\m_1\n_1}^{a_1}(x_1) F_{\m_2\n_2}^{a_2}(x_2)
F_{\m_3\n_3}^{a_3}(0) \right\} \right| \b \right\rangle, \label{double}
\ee
which must be symmetric under
$ \left( k_r , \mu_r , \nu_r , a_r \right) \leftrightarrow
\left( k_s , \mu_s , \nu_s , a_s \right) $, and
antisymmetric under $ \left( \mu_r , \nu_r \right)
\leftrightarrow \left( \nu_r , \mu_r \right) $ where $ r,s = 1,2,3$.
In order to present the expressions symmetrically we define
\be
k_3 = - k_1 - k_2.
\ee

The canonical mass dimension of ${\cal C}$ equals $ -2 $ which implies
that the only terms in the OPE which survive the double limits are
proportional to the identity operator~\footnote{As in \cite{mw1}, we assume
that
the sum of three double commutators is a renormalization group invariant
quantity.},
\be
{\cal C}^{a_1 a_2 a_3 }_{ \m_1 \n_1 \m_2 \n_2 \m_3 \n_3 } =
c^{a_1 a_2 a_3 }_{ \m_1 \n_1 \m_2 \n_2 \m_3 \n_3 }\,{\bf 1}+ \cdots,
\ee
where the remaining terms will not contribute to the final result.
The Wilson coefficients multiplying the identity operator will be such that
$[c^{a_1 \cdots}_{\mu_1 \cdots}]=({\rm mass})^{-2}$.

The coefficient function $c$ consists of a sum of terms each of
which takes the form
\be
{ \overbrace{ k \otimes \cdots \otimes k }^{n \; \rm factors} \over
\left( {\rm polynomial\; of\; degree\; } l {\rm \; in\; the\; } k_i^2
\right) } \label{wilsonc}
\ee
For the present calculation we must have $ n = 2 ( l-1 ) $.

In restricting the values of $n$ note first that all terms of the form $
k_i \cdot k_j $ can be turned into a linear combination of the $ k_i^2 $
by using $ k_1 + k_2 + k_3 = 0 $; also note that multiplying the above
expression by a dimensionless function will, at most, modify the final
result by an overall multiplicative constant, thus we can replace (for
$l > m $)
\be
{ \left( {\rm polynomial\; of\; degree\; } m {\rm \; in\; } k_i^2
\right)
\over
\left( {\rm polynomial\; of\; degree\; } l {\rm \; in\; } k_i^2
\right) }
\rightarrow
{ 1 \over
\left( {\rm polynomial\; of\; degree\; } l - m {\rm \; in\; } k_i^2
\right) }
\ee
which implies that we can ignore all contributions to $c$ containing
factors of the form $ k_i \cdot k_j $ in the numerator. Using this and the fact
that
there are six ``external'' indices $ \mu_{ 1 , 2, 3 } , \, \nu_{ 1,2,3 } $
and noting that we need include at most one $ \epsilon$ tensor, we find that we
can
restrict ourselves to $ n =0,2,4,6 $. We will consider the case $ n =
0 $ in detail, the others can be treated in the same way.

The coefficient corresponding to $ n = 0 $ in (\ref{wilsonc})
takes the form
\be
\sum_\pi \tau_{ \mu_{ \pi 1 } \nu_{ \pi 1 } \mu_{ \pi 2 } \nu_{ \pi 2 }
\mu_{ \pi 3 } \nu_{ \pi 3 } } u^{ a_{ \pi 1 } a_{ \pi 2 } a_{ \pi 3 }
} \left( \sum_r x_r k_{\pi r}^2 \right)^{-1} \label{nzeroterm}
\ee
where $ \pi $ denotes a permutation of $1,2,3$; the summation is over
the 3! such permutations. The tensor $ \tau $ is
constructed out of the metric and the $ \epsilon $ tensor. Since the
tensor $u$ takes values on a Lie algebra, its general expression will be
of the form
\be
u^{ a b c }= u_1 f^{ a b c } + u_2 d^{ a b c}, \label{defofu}
\ee
 where $f$ denotes the
(completely antisymmetric) group structure constants and $ d^{ a b c} $
denotes the completely symmetric object tr$ T^a \{ T^b , T^c \} $ ($ T^a $
denote the group generators).

Consider now the limit
$ \blim {k_r}{k_s} $, abbreviated $ \blim rs $, and let $u$ be the
(unique) index $ \not= r,\, s$. The polynomial in the denominator can be
written
\def\xt{ \tilde x }
\be
\left( \xt_r + \xt_u \right) k_r^2 + \left( \xt_s + \xt_u \right) k_s^2 +
2 \xt_u \, k_r \cdot k_s; \qquad ( u \not = r,s )
\ee
where $ \xt_r = x_{ \pi^{-1} r } $ and where we used
$ \sum_r x_r k_{\pi r}^2 = \sum_r x_{ {\pi^{-1} r} } k_r^2 $.
Then we have
\be
\blim rs {1 \over \sum_r \xt_r k_r^2 } = {1 \over 2 \xt_u } \, \delta_{
\xt_s + \xt_u }
\ee
where $ \delta_{x+x'} $ denotes the Kronecker delta. The above expression
implies
\be
\left( \blim12 + \blim23 + \blim31 \right) {1 \over \sum_r \xt_r k_r^2 }
= { 1 \over 2 } \left[
{ \delta_{ \xt_2 + \xt_3 } \over \xt_3 } +
{ \delta_{ \xt_3 + \xt_1 } \over \xt_1 } +
{ \delta_{ \xt_1 + \xt_2 } \over \xt_2 } \right] = A ( \xt_1 , \xt_2 ,
\xt_3 ) \label{zeronlimit}
\ee
$ A $ is a completely antisymmetric function of the $ \xt $ and so
\be
A ( \xt_1 , \xt_2 , \xt_3 ) = \nu_\pi \; A \left( x_1 , x_2 , x_3 \right)
\ee
where $ \nu_\pi $ denotes
the signature of the permutation $ \pi $.

Note that $A$ vanishes unless the sum of two of the parameters $ \tilde
x $ is zero which is not usually realized,
within perturbation theory. This can be understood by noting
that expression (\ref{nzeroterm}) will present poles whenever $ A \not=0
$ and $ k_i^0 \propto k_j^0 $ (neglecting the spatial components since $
k^0_i \rightarrow \infty $). In the vicinity of such poles the Wilson
coefficient behaves as $1/( k_i^2 - \eta^2 k_j^2 )$ with $ \eta $ a real
constant depending on the $ x_r $. Such behavior is rarely
generated within perturbation theory \cite{smatrix}; the high-energy
behavior of the triangle graph is, however, an
exception~\cite{triangle}.

The term under consideration
then contributes to the sum of the three double limits the quantity
\be
A \left( x_1 , x_2 , x_3 \right) \sum_\pi \nu_\pi \,
\tau_{ \mu_{ \pi 1 } \nu_{ \pi 1 } \mu_{ \pi 2 } \nu_{ \pi 2 }
\mu_{ \pi 3 } \nu_{ \pi 3 } } u^{ a_{ \pi 1 } a_{ \pi 2 } a_{ \pi 3 }
}
\ee
For the case of interest $ \mu_r = 0 $ and $ \nu_r = i_r \not= 0 $
whence, of
all possible contributions to $ \tau $, only the term containing the $
\epsilon $ tensor contributes. This leads to a term proportional to the
three-dimensional antisymmetric tensor,
\be
\tau_{ 0 \, i_{ \pi 1 } 0 \, i_{ \pi 2 } 0 \,i_{ \pi 3 } } = \tilde \tau
\epsilon_{ i_{ \pi 1 } i_{ \pi 2 } i_{ \pi 3 } } = \tilde \tau \nu_\pi \,
\epsilon_{ i_1 i_2 i_3 }
\ee
for some constant $ \tilde \tau $. The contribution to the limits then becomes
\be
A \, \tilde \tau \, \epsilon_{ i_1 i_2 i_3 } \sum_\pi u^{ a_{ \pi 1 } a_{ \pi 2
}
a_{ \pi 3 } } = \bar \tau \epsilon_{ i_1 i_2 i_3 } d^{ a_1 a_2 a_3 }
\ee
where we used the expression (\ref{defofu}) and $ \bar \tau = 6 u_2 \tilde
\tau \, A ( x_1 , x_2 , x_3 ) $. The terms containing $ f^{a b c } $ in
(\ref{defofu}) do not contribute.

The other cases, $ n = 1,\, 2,\, 3$, although more involved, yield the
same type of expressions. Collecting all results we obtain
\be
{\cal J} \left[ E^{a_1}_{i_1}(\vec x) , E^{a_2}_{i_2}(\vec y) ,
E^{a_3}_{i_3}(\vec z) \right] = \bar c \,\e_{ i_1 i_2 i_3 }
{\rm Tr}\lbrace T^{a_1} ,T^{a_2} \rbrace T^{a_3}\,\d^3(\vec x - \vec
y)\d^3(\vec x - \vec z). \label{ejacobiid}
\ee
where $\bar c$ is an undetermined constant. We note that this result
also satisfies the closure relation (\ref{closurerel}). A similar
expression
was obtained in Ref.~\cite{jo}; we will compare the present approach
with the one followed in this reference in section 4.

\bs

For the two-dimensional case only the terms containing $ F_{ \alpha
\beta }^b $ in the OPE contribute to the double limits. The coefficient
functions take the same form as in (\ref{wilsonc}) where now we have
$ n \le 4 $. After a short calculation we obtain
\be
{\cal J} \left[ E^{a_1}(\vec x) , E^{a_2}(\vec y) ,
E^{a_3}(\vec z) \right] = {\bar c}' \, u^{a_1 a_2 a_3 b} E^b \,\d^3(\vec x -
\vec
y)\d^3(\vec x - \vec z); \qquad (1+1 \ \rm{ dim.}) \label{ejacobiid2d}
\ee
where $u^{a b c d} $ is antisymmetric in its first three indices and
must be constructed out of traces of products of generators,
\be
u^{a_1 a_2 a_3 b} = i
\left[ f_{ a_1 a_2 c } d_{ c a_3 b} +
f_{ a_2 a_3 c } d_{ c a_1 b} +
f_{ a_3 a_1 c } d_{ c a_2 b} \right] \label{u4index}
\ee
Using this expression (\ref{ejacobiid2d}) is seen to satisfy
(\ref{closurerel}).

\begin{center}
\subsection*{4. The 3-cocycle in current algebra}
\end{center}

We now follow the above procedure to study the Jacobi
identity for three non-Abelian charges. We start from a gauge theory
with anti-hermitian generators $ \{ T^a \} $ and assume that a set of
current operators $ J_\mu^a $ can be defined (we will not need to
specify the chirality properties of these currents).
We then consider the operator
\be
{\cal C}^{abc}_{\m\n\r}(k_1, k_2) = \int d^4 x d^4 y
e^{i(k_1 x+ k_2 y)} T^*\left\{ J^{a_1}_{\mu_1}(x) J^{a_2}_{\mu_2}(y)
J^{a_3}_{\mu_3}(0) \right\} , \label{currents}
\ee
which is symmetric under any permutation $ ( k_s ,\, \mu_s ,\, a_s )
\rightarrow ( k_r ,\, \mu_r ,\, a_r ) $ for $ r,s = 1,2,3 $. As in the
previous section we consider first the four-dimensional case and then
briefly state the results for the two-dimensional theory.

We now expand $ {\cal C}^{abc}_{\m\n\r} $ in a series of local operators.
The terms that
will contribute to the double limits are proportional to the operators
${\bf1}$, $ F_{\mu\nu}^a $, $ \tilde F_{\mu\nu}^a $,
$ J_\alpha^b $, $ \left( D_\alpha F_{ \beta \gamma } \right)^b $,
and $ \left( D_\alpha \tilde F_{ \beta \gamma } \right)^b $. The
general expressions for arbitrary values of the indices are quite
involved and not very illuminating; we will therefore consider only
two cases: the terms proportional to the unit operator (corresponding to
the vacuum expectation value of $ \jcal $), and the case $ \mu_i =0$ which
can lead to violations of the Jacobi identity in the global algebra
generated by the charges.

\bigskip\subsubsection*{4.1 Terms proportional to {\bf1}}

We consider the Wilson coefficient associated with the unit operator
first. Using the same arguments as for the previous section we conclude
that the Wilson coefficient should take the same form as in
(\ref{wilsonc}) with $ l=1 , n=3 $, explicitly
\be
c_{\bf1}=\sum_{ r, s, t, \pi }
\tau^{rst}_{ \mu_{\pi1} \mu_{\pi2} \mu_{\pi3} \alpha \beta \gamma }\,
k_{ \pi r}^\alpha k_{\pi s}^\beta k_{\pi t}^\gamma \, u_{r s t}^{ a_{\pi1}
a_{ \pi2} a_{ \pi3} } \left( \sum_u x^{ r s t }_u k_{ \pi u}
\right)^{-1}
\ee

The evaluation of the three double limits is essentially the same as
for the previous case and we will omit the details. We obtain
\begin{eqnarray}
\left( \blim12 + \blim23 + \blim31 \right) c_{\bf1}  &=& \sum
\nu_\pi \,\bar\tau^{d ; rst}_{ \mu_{\pi1} \mu_{\pi2} \mu_{\pi3} i j n }\,
k_{ \pi r}^i k_{\pi s}^j k_{\pi t}^n \, d_{ a_1 a_2 a_3 } \\ \nonumber
&& + \sum
\bar\tau^{f ; rst}_{ \mu_{\pi1} \mu_{\pi2} \mu_{\pi3} i j n }\,
k_{ \pi r}^i k_{\pi s}^j k_{\pi t}^n \, f_{ a_1 a_2 a_3 } \\ \nonumber
\end{eqnarray}
where $ \nu_\pi $ is the signature of the permutation $ \pi $.

The tensors $ \bar \tau^d $ and $ \bar \tau^f $ must be constructed out
of the metric and the $ \epsilon $ tensor. This implies that the result
vanishes when $ \mu_i = 0 $, {\it i.e.} $ \left\langle 0 \left| \jcal
\left[ J_0^a J_0^b J_0^c \right] \right| 0 \right\rangle = 0 $,
as verified by explicit perturbative
calculations~\cite{3cocy3}. If we consider the case $ \mu_i = j_i  \not=
0 $ we obtain terms $ \sim k_r^{j_1} k_s^{j_2} k_t^{j_3}
$ and $ \sim k_r^{j_1} \delta_{ j_1 j_2 } \vec k_s^2 $ proportional to $
f_{ a_1 a_2 a_3 } $ which also agree with the results obtained
perturbation theory~\cite{3cocy3}.

\bigskip\subsubsection*{4.2 Jacobi identity for the global current algebra}

In studying violations of the Jacobi identity in the algebra of
the non-Abelian charges we define
\be
Q^a = \int d^3 \vec x \, J_0^a
\ee
so that when calculating $ { \cal J } \left[ Q^{a_1} , Q^{a_2} , Q^{a_3}
\right] $ (see (\ref{jacobi})) we need only consider only the case
$ \mu_{1,2,3}= 0 $.

In the previous subsection we showed that there are no contributions to
the operator $ \jcal[ J_0^{a_1} J_0^{a_2} J_0^{a_3}  ] $ proportional to
the unit operator.
The contributions proportional to the operators $ F $ and $ \tilde F$
have Wilson coefficients of the same form as in  (\ref{wilsonc})
with $ n = 2 l-1, \; l=1,2,3$. When $ \mu_i = 0 $ the various terms
resulting form the three double limits are  proportional to $
\vec k_r \cdot \vec E^b $ or $ { \vec k_r \cdot B^b } $, ($r=1,2,3$).
Thus they will
not contribute to the global algebra (for which we set $ \vec k_r = 0 $).

Next we consider the Wilson coefficients associated with $ J_\alpha^b $.
These again take the form (\ref{wilsonc}) with $ n = 2( l-1 ) , \;
l=1,2,3 $.
As an example we study the $ l=1 $ case; the explicit form of the
coefficient function is
\be
c_{J ; \mu_1 \mu_2 \mu_3 \alpha }^{ a_1 a_2 a_3 b } =
\sum_\pi \tau_{ \mu_{ \pi 1 } \mu_{ \pi 2 }
\mu_{ \pi 3 } \alpha} u^{ a_{ \pi 1 } a_{ \pi 2 } a_{ \pi 3 } b
} \left( \sum_r x_r k_{\pi r}^2 \right)^{-2}
\ee
where, as above, $ \pi $ denotes a permutation of 1,2,3 and $u^{ a b c d
} $ is constructed from the traces of four group generators with all
possible orderings. Evaluation of
the contribution to the three double limits is almost identical to the
one described above. As a result we get
\be
\left( \blim12 + \blim23 + \blim31 \right) c_{J} =
A ( x_1 , x_2 , x_3 ) \, \sum_\pi \nu_\pi \tau_{ \mu_{ \pi 1 } \mu_{ \pi 2 }
\mu_{ \pi 3 } \alpha} u^{ a_{ \pi 1 } a_{ \pi 2 } a_{ \pi 3 } b
}
\ee
with $A$ defined in (\ref{zeronlimit}).

Using then
$ \tau_{ 0 0 0 \alpha }= \bar \tau g_{ \alpha 0 } $ for some constant $
\bar \tau $, we find that the term containing $ J_\mu^a $ in
(\ref{currents}) contributes the operator
\be
\left( \bar \tau \, A \, \sum_\pi \nu_\pi u^{ a_{ \pi 1 } a_{ \pi 2 }
a_{ \pi 3 } b } \right) Q^b.
\ee
to $ \jcal[ Q_{a_1} Q_{a_2} Q_{a_3} ]$.
The $ n =2 $ and $ n=4 $ cases yield expressions of the same form.

Finally we consider the contributions proportional to the operators $
D_\alpha F_{ \beta \gamma } $. In this case the coefficients are of the
form (\ref{wilsonc}) with $ n = 2 ( l - 1 ) , \; l \le 4 $. Again
concentrating on the charge
operators we require $ \mu_{ 1 ,2 , 3 } = 0 $ and obtain, following the
procedure outlined in the previous section, that the contribution to the
sum of the three double limits is of the form
\be
\hbox{const} \int d^3 \vec x \left( D^\mu F_{\mu 0 } \right)^b
 \sum_\pi \nu_\pi u^{ a_{ \pi 1 } a_{ \pi 2 } a_{ \pi 3 } b } .
\ee
An identical procedure can be followed for $ D \tilde F $. The resulting
expressions contain $ D^\mu \tilde F_{ \mu 0 } $ and vanish by
virtue of the Bianchi identities.

Collecting the above results we conclude that
\be
{\cal J } \left[ Q^{a_1} , Q^{a_2} , Q^{a_3} \right] = \left[ \bar c_J
Q^b + \bar c_{DF}
\left( \int d^3 \vec x D^\mu F_{\mu 0 } \right)^b \right]
\sum_\pi \nu_\pi u^{ a_{ \pi 1 } a_{ \pi 2 } a_{ \pi 3 } b } .
\label{jiforQ}
\ee
for some constants $ \bar c_J $ and $ \bar c_{ DF } $.
Noting that $u$ must be constructed out of traces of generators, and
using the Jacobi identity for the generators we obtain that this tensor
takes the form (\ref{u4index}). It is easy to see that (\ref{jiforQ})
satisfies (\ref{closurerel}).

If we use the equations of motion $ D^\mu F_{ \mu \nu } = J_\nu $ and
define $ \bar c = \bar c_J + \bar c_{ DF } $ we obtain
\be
{\cal J } \left[ Q^{a_1} , Q^{a_2} , Q^{a_3} \right] = \bar c Q^b
\left[ f_{ a_1 a_2 c } d_{ c a_3 b} + f_{ a_2 a_3 c } d_{ c a_1 b} +
f_{ a_3 a_1 c } d_{ c a_2 b} \right]
\ee
For example, for $ SU(3) $ we have $ {\cal J } \left[ Q^1 , Q^2 , Q^3 \right]
= ( \bar c \sqrt{3} / 2 ) Q^8 $.

We can follow exactly the same procedure for the Gauss' identity
operators
\be
G^a =Q^a - \int d^3 \vec x \left( D^\mu F_{ \mu \nu } \right)^a
\ee
assuming that these operators close into an algebra we obtain that
the Jacobi identity is violated,
\be
{\cal J } \left[ G^{a_1} , G^{a_2} , G^{a_3} \right] = \bar c' G^b
\left[ f_{ a_1 a_2 c } d_{ c a_3 b} + f_{ a_2 a_3 c } d_{ c a_1 b} +
f_{ a_3 a_1 c } d_{ c a_2 b} \right]
\ee
for those cases where $ \bar c' \not= 0 $.
Note however that in the physical subspace, which is annihilated by the
$ G^a $, the Jacobi identity is valid (this would not be true is the
$G^a$ fail to close into an algebra under commutation).

\bigskip\subsubsection*{4.2 1+1 dimensions}

For the two-dimensional case some of the metric tensors which appear in
the Wilson coefficients can be replaced by the antisymmetric tensor $
\epsilon_{ \mu \nu } $. Expressing the results in terms of the left and
right-handed currents $ J_{ L , R } = J_0 \mp J_1 $ we obtain
\be
\left\langle 0 \left| \jcal
\left[ J_{h_1}^a J_{h_2}^b J_{h_3}^c \right] \right| 0 \right\rangle
= \sum_r \left[
c_{d;  h_1 h_2 h_3 }^r \, d_{a b c } +
c_{f;  h_1 h_2 h_3 }^r \, f_{a b c } \right]  K_r
\ee
where $ h_i = L,R $ and $ K_r $ denotes the spatial component of $ k_r
$.

Turning now to the global current algebra one can easily verify that the
terms in the OPE containing the operators $ \tilde F^a $ do not
generate violations of the Jacobi identity. In contrast, terms containing
the operators $ J_\mu^a $ do contribute. A straightforward
calculation (almost identical to the one described in the case of
$4$ dimensions) gives
\be
{\cal J } \left[ Q^{a_1} , Q^{a_2} , Q^{a_3} \right] = \left( \sum_{ h =
L,R } \bar c_h Q^b_h \right)
\left[ f_{ a_1 a_2 c } d_{ c a_3 b} + f_{ a_2 a_3 c } d_{ c a_1 b} +
f_{ a_3 a_1 c } d_{ c a_2 b} \right]
\ee
where $ Q_h^a = \int dx \, J_h^a $.

\begin{center}
\subsection*{5. Perturbative calculations.}
\end{center}

The previous results can be compared to the results obtained using
perturbation theory. For the vacuum expectation value of the operator $
\jcal [ J J J ] $ in $4$ dimensions the results are known~\cite{3cocy3}
and agree with the results obtained in sect. 4. The origin of the
non-vanishing contribution of such a graph can be traced to the peculiar
behavior of the discontinuities of the form factors for the triangle
graph~\cite{triangle}.

The situation is different when we consider the graphs contributing to
$ \jcal[ E E E ] $ calculated in section 3.

The expression for $ \bar c $ in (\ref{ejacobiid}) can be derived from
perturbation theory by obtaining the corresponding Wilson coefficients.
To this end we first note that the vertices corresponding to the
(composite) operator $ F^{ \mu \nu } $ are of the form

\font\smallitii=cmmi7 scaled\magstep2
\font\smalli=cmr8 scaled\magstep1
\setbox2=\vbox to 150pt{\epsfysize=4 truein\epsfbox[0 0 612 792]{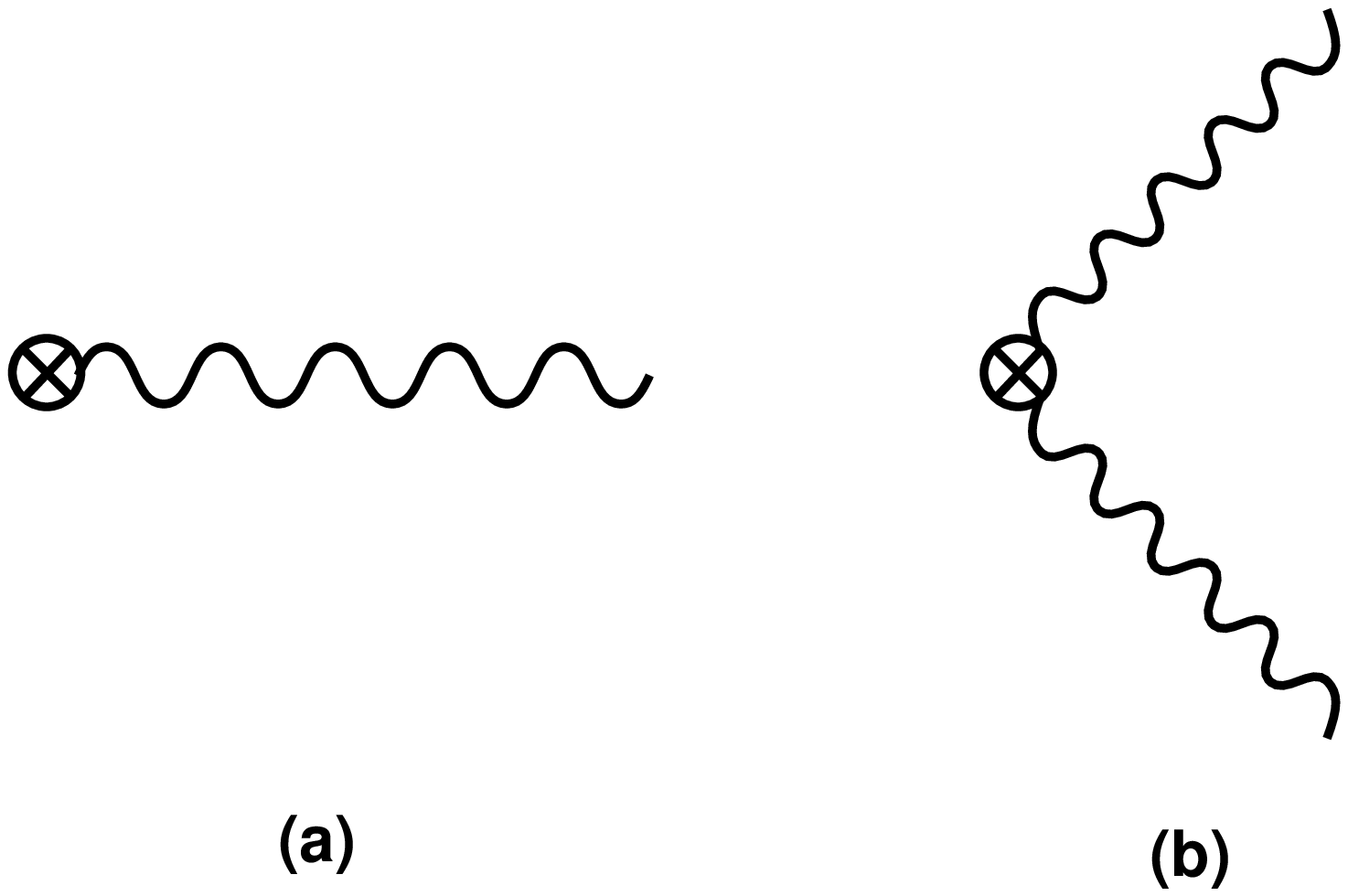}}
\centerline{\box2}

\centerline{\hfil\vbox to .3 in{\hsize 5 in
{\textindent{\smallitii Fig:{\rm\ } 1{\rm\ }}
 \global \advance \baselineskip by -10 pt
 \smalli\noindent Vertices in the composite operator
 $F_{\mu\nu}$} \vfil}\hfil}\global \advance \baselineskip by 10 pt
\bigskip

The one loop contributions are given by the graph in Fig. 2 which,
however, does not contribute to the double limits.
In fact, it is a straightforward exercise to show that any graph with one or
more
of the vertices of type (a) in Fig. 1 with one gauge boson line will not
contribute to
the three limits. This implies that the leading contributions to $ \bar
c $ are at least $ O ( g^7 ) $, where $g$ is the gauge coupling constant, and
occur at the three loop level. We will not evaluate the corresponding
graphs in this paper.

\setbox2=\vbox to 100pt{\epsfysize=4 truein\epsfbox[0 0 612 792]{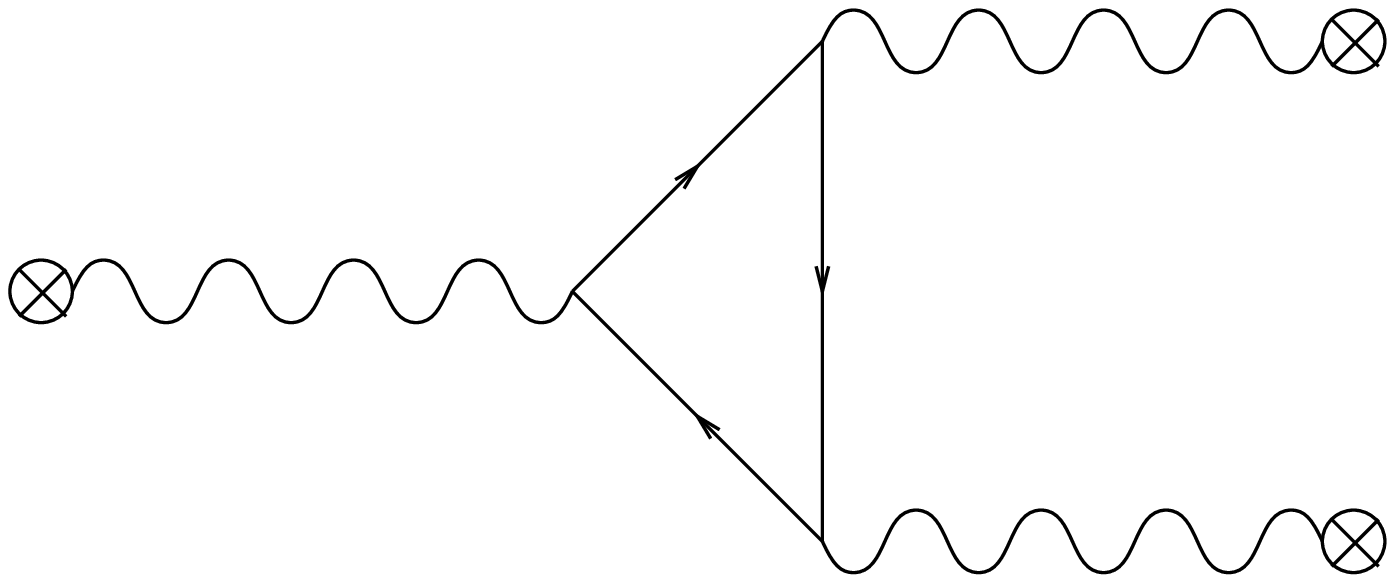}}
\centerline{\box2}

\centerline{\hfil\vbox to .3 in{\hsize 5 in
{\textindent{\smallitii Fig:{\rm\ } 2{\rm\ }}
 \global \advance \baselineskip by -10 pt
 \smalli\noindent One loop graph contributing to the Jacobi identity for
three electric fields. }\vfil}\hfil}\global \advance \baselineskip by 10 pt
\bigskip

The absence of perturbative contributions, at least at low orders,
to (\ref{ejacobiid})
contradicts the results obtained in \cite{jo} where (\ref{ejacobiid})
was obtained by first
calculating the anomalous commutator of two electric fields and then using
canonical commutation relations in the evaluation of the Jacobi operator
(\ref{jacobi}). In that calculation the
commutator of two electric fields was found to be
proportional to the gauge field, which raises questions about the gauge
invariance of the result~\footnote{A similar situation was discussed in
 \cite{mw1}. Note, however, that gauge invariance is not an issue when
considering anomalous theories.
}. In view of these problems we revisit calculation of the
commutator of two electric fields following the approach described in
 \cite{mw1}. We define \be {\cal T }^{ \mu \nu \alpha \beta } ( p ) =
\int d^4 x \, e^{ - i p \cdot x } F^{\mu \nu } ( x/2 ) \, F^{ \alpha
\beta } ( -x/2 ) \ee whose OPE takes the form $ \tau^{ \mu \nu \alpha
\beta } ( p ) {\bf1} + \cdots $ where $ \tau $ is a tensor constructed
from $p$, the metric and the $ \epsilon $ tensor; the term
proportional to the unit operator is
the only one that contributes to the BJL limit.

 The terms in $ \tau $
that generate a non-trivial BJL limit are of the form $ g^{ \mu \alpha }
p^\nu p^\beta / p^2 \pm \hbox{perms} $, and $ \epsilon^{ \mu \nu \alpha
\gamma } p_\gamma p^\beta / p^2 \pm \hbox{perms} $; where ``perms''
denotes similar terms with the indices exchanged to insure antisymmetry
under $ \mu \leftrightarrow \nu $, and $ \alpha \leftrightarrow \beta $;
and symmetry under the exchange $ ( \mu \nu ; p) \leftrightarrow ( \alpha
\beta ; -p ) $. The commutator of two electric fields is obtained from the
limit
\be \lim_{ p_0 \rightarrow \infty } p^0 \tau^{ 0 i 0 j } \ee which,
using the above expression for $ \tau $ is seen to vanish. We therefore
conclude that \be \left[ E^i_a ( \vec x/2 ) , \, E^j_a ( - \vec x/2 )
\right] = 0 \ee which disagrees with the results of \cite{jo}.

We believe that this discrepancy is due to the following. In \cite{jo}
the commutator was computed from the seagull for the current-current
commutator by using Amp\`ere's law to relate the current to $ \dot
E_i^a $. The problem with this calculation is that the seagull is not
unique, always being defined up to a covariant local contribution
 \cite{Jackiwbook}. So, if we follow the procedure described in \cite{jo}
but add to the seagull the covariant contribution
\be
\sigma_{\rm cov}^{\mu \nu}{}_{ \; ab} ( x , y ) = { i \xi \over 24 \pi^2}
{\rm Tr } \{ T^a , T^b \} \,
\epsilon^{ \mu \nu \alpha \beta } A^b_\alpha ( y ) \partial_\beta
\delta^{(4)} ( x - y )
\ee
the final result is the one in \cite{jo} multiplied by $ 1 - \xi $. The
calculation using the OPE shows that, in fact, $ \xi = 1 $.
This also implies that the
expression (\ref{ejacobiid}) cannot be derived from (\ref{jacobi}) by
evaluating some commutators canonically and others using the BJL limit.

We now consider the perturbative evaluation of the
Jacobi identity for three
currents. Following our approach we will be interested in the Wilson
coefficient for the term $ J_\alpha^b $ in the OPE of the operator $ T^*
\{ J J J \} $. The one loop contributions to the OPE are
obtained from the graphs in Fig. 3.

\setbox2=\vbox to 210pt{\epsfysize=4 truein\epsfbox[0 0 612 792]{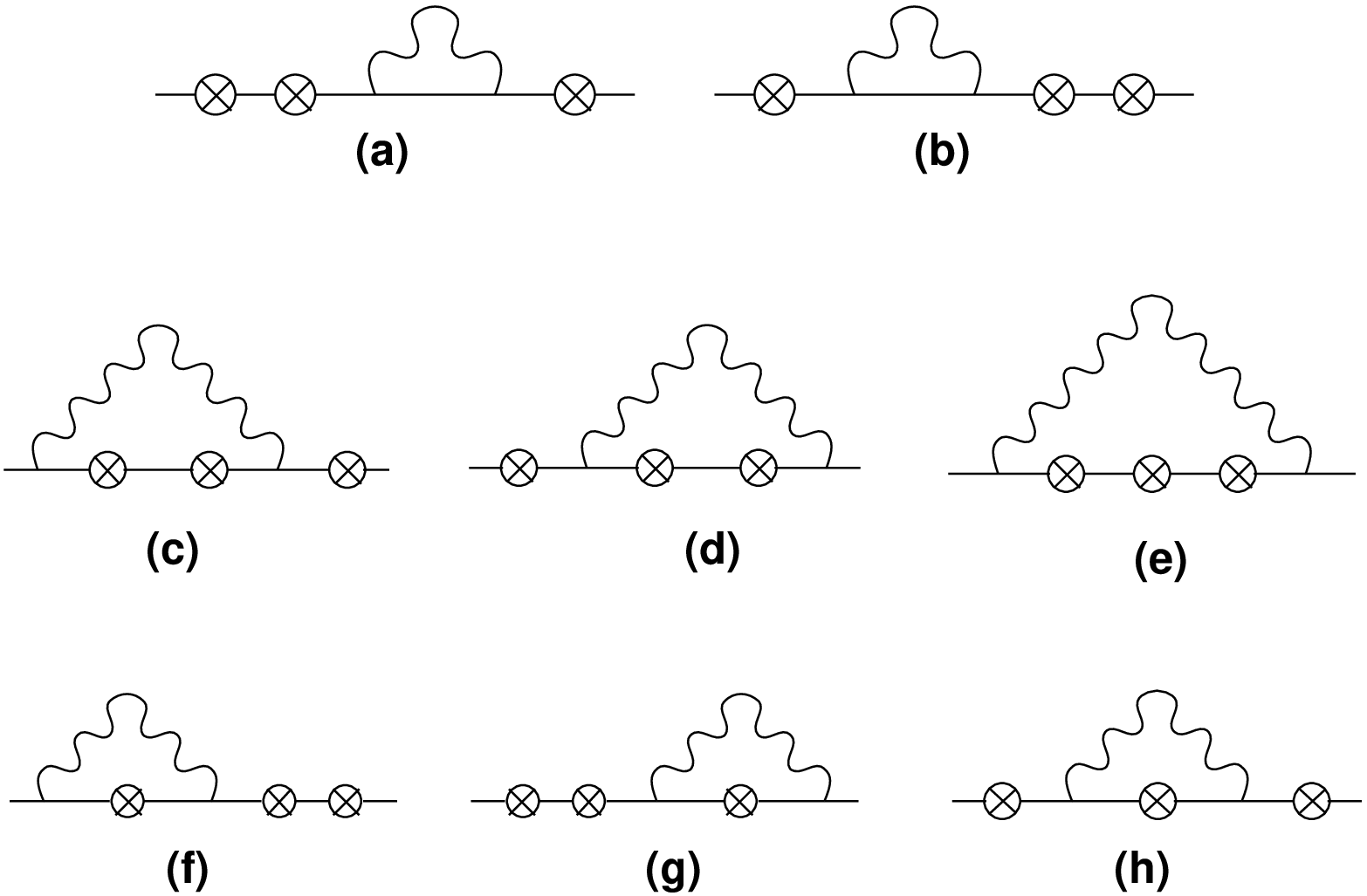}}
\centerline{\box2}

\centerline{\hfil\vbox to .3 in{\hsize 5 in
{\textindent{\smallitii Fig:{\rm\ } 3{\rm\ }}
 \global \advance \baselineskip by -10 pt
 \smalli\noindent One loop graphs contributing to the Jacobi identity for
three currents. }\vfil}\hfil}\global \advance \baselineskip by 10 pt
\bigskip

In this calculation of $ {\cal J} $
the contribution from graphs (cf. fig. 3) (a) and (b)
cancel each other; similarly graphs (c) and (d) cancel, while
the contributions of (f), (g) and (h) add up to zero.  Graph (e)
requires careful evaluation; we chose to regulate the theory using a
higher covariant derivative method~\cite{regulator}
 in the gauge-boson sector~\footnote{This
regulator induces several new vertices in the theory, but this does not
affect the present calculation.}. The propagator then
becomes (in the Feynman gauge) $ g_{ \mu \nu} / [ p^2 (1 - p^2/\Lambda^2
) ]$ and we obtain that
\be
\blim ji \left\{ \hbox{graph\ 3(e)} \right\} =
{ g^2 \over 16 \pi^2 } \left( \lambda^b \left[ \lambda^{a_j} \, , \,
\left[ \lambda^{a_k} \, , \, \lambda^{a_i} \right]
\right] \lambda^b \right) \, \ln { \Lambda^2 \over m^2 }
\ee
where $ \{ \lambda^a \} $ denote the (Hermitian) generators of the group,
$g$ the gauge coupling constant and $m$ the fermion mass. It is clear
from the above expression that the sum of the three limits vanishes by
virtue of the Jacobi identity obeyed by the group generators $ \lambda^a
$. We therefore conclude that the constant $ \bar c_J $ in
(\ref{jiforQ}) is zero to this order (see~\cite{kubo} for a related result).

\begin{center}
\subsection*{5. Conclusions}
\end{center}

We considered the simultaneous use of the operator product expansion
(OPE) and the Bjorken-Johnson-Low (BJL)
limit techniques to study double commutators
and thus look into possible violations of the Jacobi identity.
The advantages of the method are
its non-perturbative nature, the fact that all
symmetries are manifest at each stage of the calculation, and
its calculational ease. The disadvantages of the method are that
all results are determined up to unknown multiplicative constants
which could, in fact, be zero (in which case no violations of the
Jacobi identity appear). We note, however, that the vanishing of
such constants would be accidental in the sense
that it is not mandated by any symmetry of the model.

We were able to isolate cases where there cannot be
violations of the Jacobi identity. For example the vacuum expectation
value of $ \left\langle 0 \right| {\cal J } [ J_0^a , J_0^b , J_0^c ]
\left| 0 \right\rangle = 0 $ (when there is no symmetry breaking),
as discussed in section 4.1. As another
example one can consider a 4-dimensional gauge
theory with a scalar field $ \phi $; in this case
$ \jcal [ \phi , \partial_\mu \phi , A_\nu ] = 0 $ identically.

As mentioned above the method proposed provides a {\em definition} of the
Jacobi operator $ {\cal J}[ A , B , C ] $ in (\ref{jacobi}).
This definition, coincides
with the naive expression ({\it i.e.}, it vanishes)
whenever the operators $A$, $B$, $C$, and
their triple products are well defined. When regularization is needed
the expression for $ \cal J $ need not vanish. The origin of this effect
can be seen as follows: the naive expression for $ \cal J $ contains two terms
of the form $ABC$, one from the first double commutator in
(\ref{jacobi}), and one from the last double commutator.
The the first equal-time commutator, however, is evaluated by
first letting
$ t_B \rightarrow t_A $ and subsequently $ t_C \rightarrow t_A $;
the second commutator is obtained by taking
$ t_C \rightarrow t_A $ and then $ t_B \rightarrow t_A $. The two limits
need not commute leading to a non-zero contribution. This can be
interpreted as a lack of associativity, $ (AB)C \not= A(BC) $ which is
related to the presence of a three cocycle. A naive definition of $ \cal
J $ would not exhibit this feature, the cost being that the operator
products are ill defined.

As applications we considered violations of the Jacobi identity for
three chromo-electric fields as well as for three non-Abelian charges and
for three Gauss' law generators. The resulting
three-cocycles satisfy the closure relation (\ref{closurerel})
and therefore imply that the corresponding group is not associative.
The general analysis
relates the violations of the Jacobi identity to poles in the Wilson
coefficient functions at large time-like momenta. Such poles are absent
in most perturbative contributions leading to $ {\cal J} = 0 $; this is
the case for three chromo-electric fields and three current
charges. The one exception we have found corresponds to those
perturbative contributions generated by the triangle graph, which
generate non-trivial violations to the Jacobi
identity of three (space-like) currents. The form of these perturbative
results agree with that obtained using the OPE and BJL limit approach.

The fact that general considerations lead to a violation of the Jacobi
identity implies, as mentioned previously, that a well-defined
representation of such operators does not exist~\cite{carey.et.al.}. For
example, a representation for the gauge field operators cannot be extended
to include the $ E_i^a $; these objects are then to be defined in terms
of their commutators with space-time smeared operators.

It is also worth noticing that even if the
Jacobi identity fails the corresponding group can still be made
associative by an appropriate quantization of the
3-cocycle~\cite{magmon}.

\bs

\subsection*{Acknowledgments}

We would like to thank R. Jackiw for various interesting comments.
This work was supported in part through funds provided by the Department
of Energy under contract DE-FG03-92ER40837.}

\bs\bs

\vfill\eject
\ed